\documentclass[aps,preprint]{revtex4-1}
\usepackage{graphicx}
\usepackage{amsmath}
\usepackage{makeidx}
\usepackage{amsfonts}
\usepackage{amssymb}
\usepackage{mathtools}
\usepackage{chngcntr}
\usepackage{xcolor}

\begin{document}

\title{Efficiency and irreversibility of movements in a city}

\author{Indaco Biazzo}
\affiliation{Politecnico di Torino, Corso Duca degli Abruzzi 24, Torino, Italy}

\author{Abolfazl Ramezanpour}
\email{aramezanpour@gmail.com}
\affiliation{Physics Department, College of Sciences, Shiraz University, Shiraz 71454, Iran}
\affiliation{Leiden Academic Centre for Drug Research, Faculty of Mathematics and Natural Sciences, Leiden University, PO Box 9500-2300 RA Leiden, The Netherlands}

\date{\today}

\begin{abstract}
We know that maximal efficiency in physical systems is attained by reversible processes. It is then interesting to see how irreversibility affects efficiency in other systems, e.g., in a city. In this study, we focus on a cyclic process of movements (home to workplace and back to home) in a city to investigate the above question. To this end, we present a minimal model of the movements, along with plausible definitions for the efficiency and irreversibility of the process; more precisely, we take the inverse of the total travel time per number of trips for efficiency and the relative entropy of the forward and backward flow distributions for the process irreversibility. We perform numerical simulations of the model for reasonable choices of the population distribution, the mobility law, and the movement strategy. The results show that the efficiency of movements is indeed negatively correlated with the above measure of irreversibility. The structure of the network and the impact of the flows on the travel times are the main factors here that affect the time intervals of arriving to destinations and returning to origins, which are usually larger than the time interval of the departures. This in turn gives rise to diverging of the backward flows from the forward ones and results to entropy (disorder or uncertainty) production in the system. The findings of this study might be helpful in characterizing more accurately the city efficiency and in better understanding of the main working principles of these complex systems. \end{abstract}


\maketitle

\section{Introduction}\label{S0}
Constructing an effective theory (macroscopic description) of a complex system with many interacting degrees of freedom, would be very helpful for understanding the system behaviour. Thermodynamics is such an example which focuses on the appropriate macroscopic properties of a system and the relevant ways of exchanging its energy with an environment. Specifically, we know that a physical system is better to work close to a reversible process in order to achieve a higher efficiency \cite{Lavenda-book-1978}. This is expected to be somehow true for biological systems which are out of equilibrium thermodynamic systems working at a nonzero power \cite{B-book-2012}. But what about other complex systems like the stock market or a city? We expect an efficient market to be close to a maximal entropy state, where nobody can systematically beat the market to make significant returns in a long run \cite{Emarket-jfe-1998,Emarket-jep-2003,Emarket-jpm-2004}. In this paper, however, we focus on a city system to investigate the extent to which the above picture holds for the process of movements in city. We define suitable measures of efficiency and irreversibility for a cycle of movements in the system. Then, by numerical simulations of the model with real and simulated population distributions, we show that the efficiency of movements is negatively correlated with the irreversibility of the process for plausible choices of the model parameters.

The science of city is mainly devoted to the application of concepts and methods of complex systems to cities \cite{ut-nat-2010,Batty-book-2013,B-book-2016,B-nr-2019}. A city is indeed an adaptive dynamical system which grows in size and population, and consumes energy to maintain its function and structure \cite{neq-ac-2006,Batty-sci-2008,tc-spr-2009,cp-sci-2011,cg-jie-2015}. In particular, there are scaling relations connecting the macroscopic variables of a city such as population, area, energy consumption, gross domestic product, and other state variables \cite{ut-nat-2010,scaling-sci-2013,scaling-plos-2014}. These scaling (self-similar) behaviours can be reproduced and explained by some stochastic models of city formation and growth \cite{Batty-book-2013,Li-nc-2017}. A city can also be viewed as a multilayer network of interdependent networks like the communication and transport networks \cite{mt-book-2011,B-book-2016}. The structure of this multiplex network affects the system dynamics (e.g. spreading processes) and function, and so the city efficiency \cite{sf-nature-1999,sus-book-2005}. City is also a system of agents making decisions based on the model that agents construct from the available information. The difference between this (ideally maximum-entropy) model and the actual one would result to inappropriate decisions and inefficiencies. This inefficiency is usually accompanied by increasing the disorder (uncertainty) or entropy production in the system, which is an essential hallmark of irreversible processes \cite{C-pre-1999,J-cm-2011,st-rpp-2012}. 
   
\begin{figure}
\includegraphics[width=10cm]{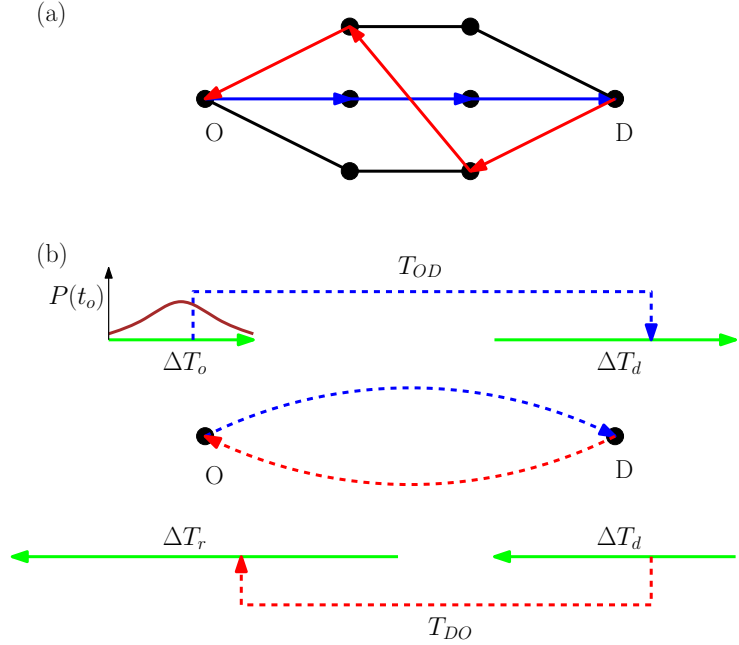} 
\caption{A cycle of movements from origins ($O$) to destinations ($D$). (a) An example of origin to destination path that differs from the destination to origin path, e.g., due to asymmetries in the travel times. (b) Schematic representation of the time intervals $\Delta T_o$, $\Delta T_d$ and $\Delta T_r$. The starting times $t_o$ of the OD trips are distributed in the time interval $\Delta T_o$ with probability distribution $P(t_o)$. In a general graph with multiple OD pairs, two trips which start from different origins at the same time may arrive at the same destination in different times. This gives rise to a destination time interval $\Delta T_d$ which is usually larger than $\Delta T_o$ and could result in backward flows which are very different from the forward flows.}\label{f1}
\end{figure}

Consider a cycle of movements from origins (home) to destinations (work) at the morning time and then back from the destinations to origins in the afternoon (see Fig.\ref{f1}). A measure of efficiency can be defined here by comparing the total travel time of the individuals with the total number of necessary travels along the edges of the network, which is expected to represent the total cost of the movements \cite{E-njp-2003}. The connectivity structure of the city and its population and work-places distribution, the flux of movements and travel strategies, are among the main factors that affect the above efficiency. Here, however, we are interested in possible relations with the irreversibility of the process. Let us assume that all the origin to destination (OD) movements start in a small time interval $\Delta T_o$. The people would arrive at the work places in a time interval $\Delta T_d$ which is expected to be larger than $\Delta T_o$. Here, $\Delta T_d$ is the time interval in which all arrivals happen. Two main reasons are at work here: the network structure and the flow dynamics. There are for example many shortest OD paths in the network which span a finite range of travel times even in the absence of intensive flows on the edges. In addition, the flows affect the travel times and even for two paths of the same length, the actual travel times could be very different because of differences in the flows. The same reasoning says that the time interval of returning back to home $\Delta T_r$ should be larger than the destination time interval. This mechanism is responsible for entropy production by increasing the uncertainty in the system and raising the cost (time or energy) we need to bring the system back to its initial state.

Now suppose that $\overrightarrow{\mathbf{f}}$ and $\overleftarrow{\mathbf{f}}$ represent the (average) flow distributions (on edges) for the forward (OD) and backward (DO) processes, respectively. The backward process is defined by reversing all the origin to destination trips. Then a measure of irreversibility can be defined by the distance or divergence of the two distributions $D(\overrightarrow{\mathbf{f}}||\overleftarrow{\mathbf{f}})$. As mentioned above, the destination to origin trips are distributed in a destination time interval $\Delta T_d$ that is usually larger than the origin time interval $\Delta T_o$. As a result, the DO travel times and flows are not necessarily the same as the OD ones. These asymmetries results to differences in the forward and backward flows and contribute to the irreversibility of the process. The above arguments suggest that a measure of irreversibility can be defined by the relative entropy of the forward and backward flow distributions or the relative entropy of the time intervals at the endpoints of the process. In the following, we present and study a minimal model of cyclic movements to make the above concepts and relations more quantitative.

\section{Models and Settings}\label{S1}
In this section, we present the main definitions and methods which are used to model the network flow dynamics.   
Consider a city of $N$ sites with local populations $\{m_a:a=1,\dots,N\}$ and total population $M=\sum_a m_a$. The connectivity graph of the city is given by $G(V,E)$ where $V$ is the set of sites and $E$ is the set of edges. We use the simple growth model introduced in Ref. \cite{Li-nc-2017} to produce reasonable population distributions for the model cities: start with an active seed of population $m_{seed}$ in the centre of a two-dimensional grid with undirected edges of unite length; A site is active if it has a nonzero population. At each time step, one node $a$ is selected with probability proportional to $m_a+c_0$ where $c_0>0$. The population at site $a$ increases by one if there is an active site $b$ close to site $a$, that is $|x_b-x_a|\le r_0$ or $|y_b-y_a|\le r_0$ for some small $r_0$. Here $(x_a,y_a)$ are the coordinates of site $a$. The above process is repeated for $10^6$ iterations, where each iteration consists of $N$ time steps. This model has been used to describe the scaling relations that are dependent on the profile of population in city \cite{Li-nc-2017}. Moreover, the qualitative behaviours of the model are not sensitive to the precise values of the parameters $m_{seed}, c_0$ and $r_0$.        

Given the population distribution $m_a$, we need a mobility law to construct the flux of movements $m_{a\to b}$ from origins $a$ to destinations $b$. Note that we do not need to have a direct connection from $a$ to $b$. There are many works that try to reproduce the observed movements in cities by a simple mobility law \cite{rad-nature-2012,Hasan-jsp-2013,Yan-intf-2014,Ren-nc-2014,mob-plos-2015}. For instance, the generalized gravity law states that $m_{a\to b}$ is proportional to $m_am_b/r_{ab}^{\alpha}$ for two sites at distance $r_{ab}$. In this study, we use the following mobility law \cite{Yan-intf-2014}:
\begin{align}\label{mab}
m_{a\to b}=m_ap_{a\to b}=m_a\frac{m_b/M(r_{ab})}{\sum_{c\neq a} m_c/M(r_{ac})},
\end{align}
where $M(r_{ab})$ is the population in the circle of radius $r_{ab}$ centred at site $b$. The ratio $m_b/M(r_{ab})$ can be interpreted as the attractiveness of site $b$ for an individual at site $a$. This model and the related generalizations are able to reproduce well the empirical data.      

Finally, the flows $F_{ab}$ of movements on edges $(ab)\in E$ are determined by a flux distribution problem that satisfies the system constraints and preferences. For instance, the flows can be obtained by minimizing the total travel time subject to the movements $m_{a\to b}$ \cite{lo-plos-2015,Serdar-nc-2016}. Here, instead, we use a more local and selfish strategy, where the movements from origin $a$ to destination $b$ go through the shortest-time path connecting the two nodes. The path is defined as the one that takes the minimum time based on the expected travel times for each edge of the network. The expected times can be obtained in a learning process using the history of the actual travel times.

Later in this section, we shall define a measure of efficiency focusing on the total travel time and the total number of trips. There are measures of transport or commuting efficiency defined in the literature addressing different aspects of the movements \cite{horner-epa-2002,sus-book-2005,newman-jstst-2006,EC-trans-2015,ceff-srep-2016,indaco-arx-2018}. The route factor and its generalizations compare the topological distances in the network with the geometrical distances \cite{newman-jstst-2006,ceff-srep-2016}. The excess commuting index on the other hand concerns with distribution of home and work places and compares the actual commuting distances with a theoretical optimal one \cite{horner-epa-2002,EC-trans-2015}. Finally, the accessibility of a city can be quantified by the velocity and sociability scores defined in \cite{indaco-arx-2018}. Each of these measures focuses on some structural or dynamical properties of the network and the commuting process. In this study, we are specifically interested in the efficiency of the process of movements concerning the travel times and the number of necessary trips (cost of travels).

\subsection{The movement process}\label{S11}
We are interested in a cycle of movements from origins to destinations and back to the origins. This is the basic motif of movement patterns in a city \cite{hmobility-nph-2010,sp-mpc-2012,motif-rci-2013,motif-nc-2017}. Let us assume that we are given the population distribution $m_a$ and the fluxes $m_{a\to b}$. Then, a cycle of the movement process is defined as follows: 

\begin{itemize}

\item The starting times of the OD trips are distributed (with a given probability measure) in a time interval $\Delta T_o$.  

\item The transport services run at time intervals $\Delta t=1$ to carry the passengers in $n_o=\lceil\Delta T_o/\Delta t\rceil$ time steps. 

\item We obtain the flows $F_{ab}(t)$ at each time step $t$ using a flow dynamics. Here $F_{ab}(t)$ is the number of people moving on edge $(ab)$ in time step $t$. A simple strategy is to choose the shortest paths according to the expected times $\tilde{t}_{ab}$, which are estimated from the previous cycles. For the initial cycle $\tilde{t}_{ab}(0)=t_{ab}(0)$, where the $t_{ab}(0)$ are the travel times for free lines.   

\item Given the flows, then the actual travel times are obtained from 
\begin{align}\label{tab}
t_{ab}(F_{ab})=t_{ab}(0)\left(1+g(\frac{F_{ab}}{F_{ab}(0)})^{\mu}\right),
\end{align}
where $g (F_{ab}/F_{ab}(0))^{\mu}$ is to model the influence of flows on the travel times \cite{lc-trans-1976,lc-trans-2011}. The nonnegative parameters $g$ and $\mu$ control the above effect. Here $F_{ab}(0)$ is a measure of the line capacity. Note that in general $t_{ba}(0)\neq t_{ab}(0)$ and $F_{ba}(0)\neq F_{ab}(0)$, for example, because of structural asymmetries.

\item the passengers return to their origin after spending time $T_w$ at their destinations. Thus, the return times are distributed in the time interval $\Delta T_r$ according to the arrival times.   

\end{itemize}

Figure \ref{f1} (panels a and b) gives an illustration of the above process with the associated time intervals for the trips from one origin to a destination.  The relevant quantities here are the total travel time and the total number of active transport services (the cost or number of trips):
\begin{align}
T &=\sum_t \sum_{(ab)}F_{ab}(t)t_{ab}(t),\\
C &=\sum_t \sum_{(ab)}\mathbb{I}(F_{ab}(t)>0).
\end{align}
Here $\sum_{(ab)}$ denotes a sum over all directed edges of the connectivity graph $G$. 
The indicator function $\mathbb{I}(.)$ is one if the enclosed condition is satisfied, otherwise it is zero. 
Here we assume that only one service runs on directed edge $(ab)$ if there exist at least one individual that has to travel along the edge. In a more realistic model, the services could have limited capacities and the number of services would depend on the number of passengers.       

Given the travel time per person $\tau=T/M$ and the cost per person $\sigma=C/M$, we define the total efficiency  $\eta=\eta_{OD}+\eta_{DO}$, with
\begin{align} 
\eta_{OD}=\frac{1/\tau_{OD}}{\sigma_{OD}},\hskip1cm
\eta_{DO}=\frac{1/\tau_{DO}}{\sigma_{DO}}.
\end{align}
Note that both the number of trips and the travel times are minimized if: (i) the OD and DO trips occur in one time step ($\Delta T_o,\Delta T_d \simeq \Delta t $) and (ii) the expected travel times $\tilde{t}_{ab}$ are close to the actual travel times ($g\to 0$). 

To define a measure of irreversibility, we first define the average forward and backward distributions 
\begin{align}
\overrightarrow{f_{ab}} &=\sum_{t \in \Delta T_o} w_{od}(t) \frac{F_{ab}(t)}{F(t)},\\
\overleftarrow{f_{ab}} &=\sum_{t \in \Delta T_d} w_{do}(t) \frac{F_{ba}(t)}{F(t)}.
\end{align}
Here $w_{od}(t)$ and $w_{do}(t)$ are the fraction of OD and DO movements at time step $t$, respectively. More precisely, $w_{od}(t)=(\sum_a\sum_b m_{a\to b}(t))/M$, and recall that starting time of the movements $m_{a\to b}$ are distributed in $\Delta T_o$ such that $m_{a\to b}=\sum_{t \in \Delta T_o}m_{a\to b}(t)$. Similarly we define the fractions $w_{do}(t)$ for the DO trips. The normalization factors are $F(t)=\sum_{(ab)}F_{ab}(t)$. Note that the backward flow $\overleftarrow{f_{ab}}$ on directed edge $(ab)$ is defined by the flows $F_{ba}$ on the edge $(ba)$ for the DO trips.  
Then, the Kullback-Leibler (KL) divergence or the relative entropy of the two probability distributions is given by 
\begin{align}
D_{KL}(\overrightarrow{\mathbf{f}}||\overleftarrow{\mathbf{f}}) &=-\sum_{(ab)} \overrightarrow{f_{ab}} \ln (\overleftarrow{f_{ab}}/\overrightarrow{f_{ab}}).
\end{align}
The KL divergence is nonnegative and it is zero only when the two distributions are the same. 
  
Another measure of entropy production in the process can be defined by considering the expansion of the time intervals $\Delta T_d$ and $\Delta T_r$ with respect to the $\Delta T_o$. To quantify this we define the relative entropy of the time intervals $\Delta S_T=\Delta S_{OD}+\Delta S_{DO}$, where 
\begin{align}
\Delta S_{OD}= \ln(n_d)-\ln(n_o),\\
\Delta S_{DO}= \ln(n_r)-\ln(n_d).
\end{align}
The number of time steps in each interval is given by $n_o=\lceil\Delta T_o/\Delta t\rceil, n_d=\lceil\Delta T_d/\Delta t\rceil$, and $n_r=\lceil\Delta T_r/\Delta t\rceil$.

\section{Results}\label{S2}
We take a two-dimensional grid of $N=L\times L$ sites for $G(V,E)$ with connectivity $z=4$ and links of length one. The population distribution ($m_a$) is constructed by simulation of the growth model described in Sec. \ref{S1} with parameters $m_{seed}=1, c_0=1, r_0=1$. For a real city, the network structure $G$ and population distribution are provided by the available data from \cite{popGridEurostat,sedacV4} (see Appendix Fig. \ref{SM0} for an example). The OD mobilities are obtained by Eq. \ref{mab}. Given the expected travel times $\tilde{t}_{ab}$, the flows $F_{ab}$ are determined by the shortest path (in time) strategy. We shall assume that the starting time of the OD trips in the time interval $\Delta T_o$ obeys a centred Gaussian distribution of standard deviation $\Delta T_o/3$. The actual travel times are computed by Eq. \ref{tab} with $F_{ab}(0)=F_{ba}(0)=M/(2|E|)$. We also assume that $t_{ab}(0)=t_{ba}(0)=1$ for all directed edges in $G$. Therefore, there is no structural asymmetry in the model. We consider a learning process in which the expected travel times are updated by using the information about the actual travel times in the previous cycle. More precisely, for cycle $n$ we take $\tilde{t}_{ab}(n)=\lambda t_{ab}(n-1)+(1-\lambda)\tilde{t}_{ab}(n-1)$, with $\lambda=1/2$ as a damping parameter and $\tilde{t}_{ab}(0)=t_{ab}(0)$. This means that the expected travel times for the next round are the average of the actual and expected times in the previous round. In other words, we are trying to find a good estimation of the travel times by slowly correcting the expected values according to the new observations. We repeat the cycle for $n_c=20$ times and report the results at the end of this process.

\begin{figure}
\includegraphics[width=16cm]{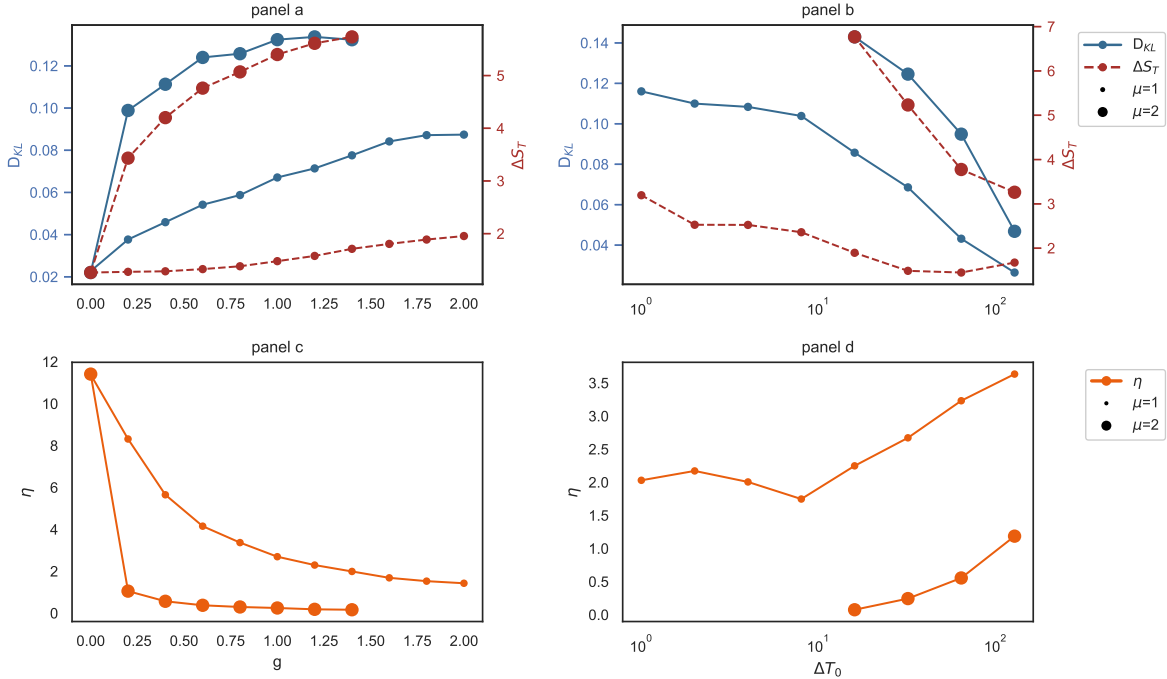} 
\caption{Variation of the average efficiency and relative entropies with $g$ and $\Delta T_o$. Panels (a) and (c): the behaviour when only $g$ changes with $\mu$ and $\Delta T_o=32$ fixed. Panels (b) and (d): the behaviour when only $\Delta T_o$ changes with $\mu$ and $g=1$ fixed. The average is taken over $100$ realizations of population distribution and movements (with learning) on a two-dimensional grid of size $N=50\times 50$. The errorbars are about the size of the larger points.}\label{f2}
\end{figure}

\begin{figure}
\includegraphics[width=16cm]{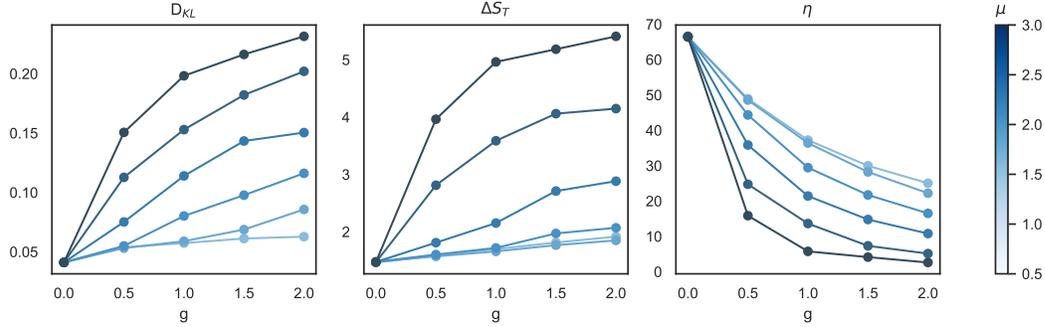} 
\caption{The average behaviour of the efficiency ($\eta$) and the relative entropies ($D_{KL}, \Delta S_T$) in $20$ real cities (the core parts). The data are obtained by numerical simulation of the movements after $20$ learning cycles using the population distributions of the cities. The average is taken over the cities for $\Delta T_o=32$. Each curve shows the behaviour for a given value of $\mu \in (0,3)$.}\label{f3}
\end{figure}

We first check the behaviour of the proposed observables ($D_{KL},\Delta S_T,\eta$) with the parameters of the model ($g, \mu, \Delta T_o$).  Two observables measure the irreversibility of the process: the KL divergence ($D_{KL}$) and the relative entropy ($\Delta S_T$). The KL divergence is divided by $\ln(2|E|)$ to be able to compare it for different city sizes. The third observable measures the efficiency of the process ($\eta$). The parameters of the model that we consider are the time window of starting the OD trips ( $\Delta T_o$) and the two variables, $\mu$ and $g$, which control the capacity of the lines (the influence of flows on the travel times). In Fig. \ref{f2} the results obtained by the simulated population distributions are shown. The $D_{KL}$ and $\Delta S_T$ increase with the parameter $g$ for the given values of $\mu=1,2$ (Fig. \ref{f2}, panel a). Instead, the same two quantities decrease when $\Delta T_o$ increases (Fig. \ref{f2},  panel b). In our interpretation, this means that the entropy production or irreversibility of the process increases when the line capacity decreases, and it decreases by enlarging the time window of the OD trips ($\Delta T_o$).  Instead, the efficiency of the movement process ($\eta$) decreases when the capacity of the lines decreases (Fig.\ref{f2}, panel c) and it increases when $\Delta T_o$ is widened (Fig. \ref{f2}, panel d). The same dependence of the quantities $D_{KL},\Delta S_T,\eta$ on the parameters $g, \mu$ is observed when the population distributions of $20$ real cities is used (Fig. \ref{f3}).

\begin{figure}
\includegraphics[width=14cm]{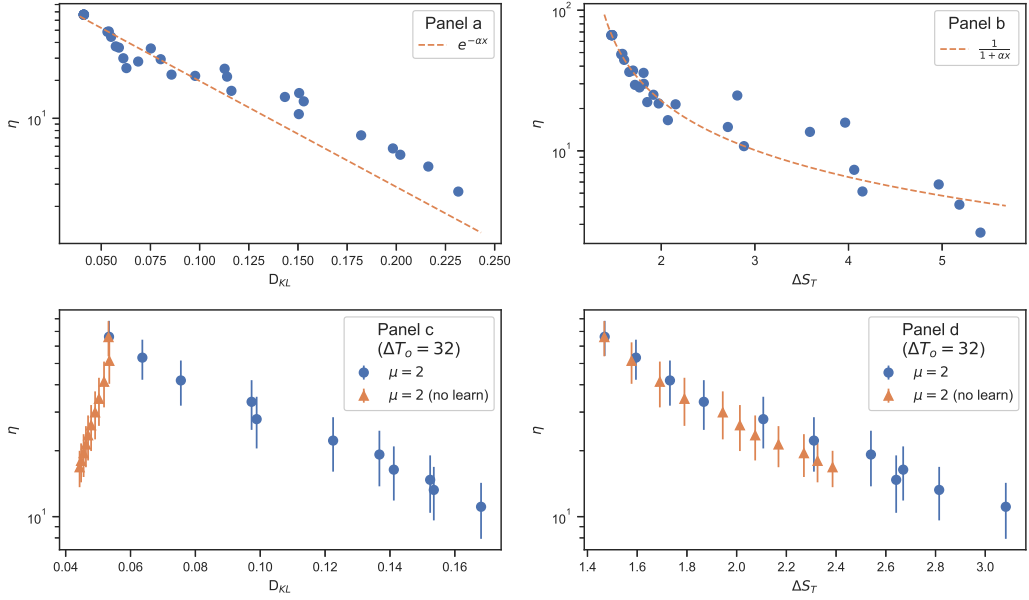} 
\caption{The average efficiency ($\eta$) vs the relative entropies ($D_{KL}, \Delta S_T$) for $20$ real cities (the core parts). Panels (a) and (b): the behaviour for different values of $g \in (0,2)$ and $\mu \in (0,3)$ for a given $\Delta T_o=32$. Panels (c) and (d): the behaviour when only $g$ changes with $\mu$ and $\Delta T_o$ fixed.}\label{f4}
\end{figure}

Figure \ref{f4} displays the dependence of $\eta$ on the $D_{KL}$ (panel a) and $\Delta S_T$ (panel b) when both the parameters $g \in (0,2)$ and $\mu \in (0,3)$ are varied for a fixed $\Delta T_o=32$. The observed behaviours, which are obtained from real population distributions, can well be described by an exponential relation $\eta \propto \exp(-\beta D_{KL})$, with exponent $\beta=19.3 \pm 1.8$. Similar behaviour is also observed with the simulated population distributions (Appendix Figs. \ref{SM1}, \ref{SM2}, and \ref{SM3}). For comparison, in Fig. \ref{f4} (panel c and panel d) we also show the results obtained with no learning, that is without any knowledge of the actual travel times in the previous cycles. Note that the learning process dose not necessarily increase the system efficiency $\eta$, because the aim of learning here is just to find the shortest (time) path. Moreover, as the figure shows, the learning process considerably changes the behaviour of the efficiency with the KL divergence. The relation with the relative entropy $\Delta S_T$, by contrast, does not qualitatively change by the learning process. The latter solely measures the changes in the size of the time intervals $\Delta T_d$ and $\Delta T_r$ which usually grow by increasing $g$ or $\Delta T_o$. On the other hand, the KL divergence is affected by both the size of the time intervals and the distribution of the arrival times in these intervals. The destination time interval $\Delta T_d$ plays a central role in this study; the network structure and the impact of the forward flows on the travel times (Eq. \ref{tab}) usually give rise to a large $\Delta T_d$ (larger than $\Delta T_o$). An extreme example is the case that all the OD trips start at the same time. And the size of $\Delta T_d$ directly affects the divergence of the backward flows from the forward ones. Here, both the forward and backward travel times $T_{OD},T_{DO}$ are expected to increase with $\Delta T_d$ (see Appendix Fig. \ref{SM4}).

Finally, we studied the cumulative distribution of the simulated OD times $T_{OD}$, the normalized flows $f_{ab}$, and the destination time intervals (across the sites), for the core parts of the cities (Appendix Fig. \ref{SM5}). Interestingly, here we observe a tendency to exhibit scale free behaviours by introducing the impact of the flows on the travel times. Note that distribution of the arrival times in $\Delta T_d$ is by definition similar to that of $T_{OD}$. Moreover, distribution of the actual travel times $t_{ab}$ is related to that of flows $f_{ab}$ after Eq. \ref{tab}; if $P(f_{ab}) \propto f_{ab}^{-\alpha}$ then one gets $P(t_{ab}) \propto t_{ab}^{-(1+(\alpha-1)/\mu)}$ for $t_{ab} \gg t_{ab}(0)$. For instance, we find $\alpha \simeq 2$ and $P(T_{OD}) \propto T_{OD}^{-\gamma}$ ($T_{OD} \gg \Delta t$) with $\gamma \simeq 3/2$ for the city Prague when $g=1, \mu=3, \Delta T_o=32$.

\section{Conclusion}\label{S3}
We observed that reasonable definitions of efficiency and irreversibility are negatively correlated in a  plausible model of movements in a city. It means that by reducing the process irreversibility one can indirectly enhance the movement efficiency. For the numerical simulation of the process, we used models that try to reproduce the main features of actual population distributions and mobility fluxes. We also used real population distributions of some real cities to compute the associated efficiency and irreversibility from the above model of movements. An empirical estimation of these quantities however needs more detailed information about the forward and backward flows, the travel times and the number of necessary trips. 

One should see how much the results of this study are robust to change in definitions of the efficiency and irreversibility. Specifically, other measures of entropy production can be studied within the framework of stochastic thermodynamics \cite{st-rpp-2012}. Also, it would be interesting to see how these quantities are related to the system criticality and predictability \cite{critical-dd-1999,predict-sci-2010,predict-jsta-2013}. The measures we introduced here are suited for the process of movements in the city. One could have such measures of efficiency and entropy production for other processes happening in a city, and so for the whole city. As already mentioned, the main task here is to find out if the negative correlation between the efficiency and irreversibility is a working principle of the cities.

\acknowledgments AR is grateful to A. Montakhab and F. Shahbazi for helpful discussions and constructive suggestions. This work has been supported by the SmartData@PoliTO center on Big Data and Data Science.

\appendix
\setcounter{figure}{0} 
\counterwithin{figure}{section}

\section{Supplemental Figures}
Here we report the results which mentioned in the main text but the reader is referred to the Supplemental Material for the figures.   

We take a two-dimensional grid of $N=L\times L$ sites for $G(V,E)$ with connectivity $z=4$ and links of length one. The population distribution ($m_a$) is constructed by simulation of the growth model described in the main text (Sec. II) \cite{Li-nc-2017} with parameters $m_{seed}=1, c_0=1, r_0=1$. For a real city, the network structure $G$ and population distribution are provided by the available data from \cite{popGridEurostat,sedacV4} (see Fig. \ref{SM0} for an example). The OD mobilities are obtained from \cite{Yan-intf-2014}:
\begin{align}
m_{a\to b}=m_ap_{a\to b}=m_a\frac{m_b/M(r_{ab})}{\sum_{c\neq a} m_c/M(r_{ac})}.
\end{align}
Given the expected travel times $\tilde{t}_{ab}$, the flows $F_{ab}$ are determined by the shortest path (in time) strategy. We shall assume that the starting time of the OD trips in the time interval $\Delta T_o$ obeys a centred Gaussian distribution of standard deviation $\Delta T_o/3$. The actual travel times are computed by
\begin{align}
t_{ab}(F_{ab})=t_{ab}(0)\left(1+g(\frac{F_{ab}}{F_{ab}(0)})^{\mu}\right),
\end{align}
with $F_{ab}(0)=F_{ba}(0)=M/(2|E|)$. We also assume that $t_{ab}(0)=t_{ba}(0)=1$ for all directed edges in $G$. Therefore, there is no structural asymmetry in the model. We consider a learning process in which the expected travel times are updated by using the information about the actual travel times in the previous cycle. More precisely, for cycle $n$ we take $\tilde{t}_{ab}(n)=\lambda t_{ab}(n-1)+(1-\lambda)\tilde{t}_{ab}(n-1)$, with $\lambda=1/2$ as a damping parameter and $\tilde{t}_{ab}(0)=t_{ab}(0)$. We repeat the cycle for $n_c=20$ times and report the results at the end of this process.

\subsection{Simulated population distributions}
\begin{figure}
\includegraphics[width=16cm]{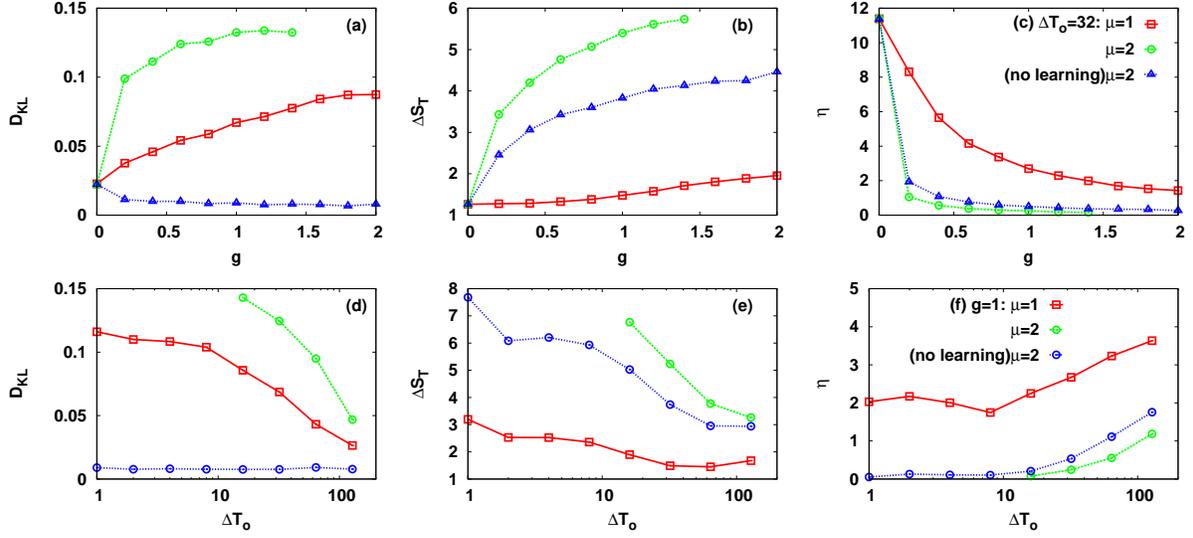} 
\caption{Variation of the average efficiency and relative entropies with $g$ and $\Delta T_o$. Top: the behaviour when only $g$ changes with $\mu$ and $\Delta T_o$ fixed. Bottom: the behaviour when only $\Delta T_o$ changes with $\mu$ and $g$ fixed. The average is taken over $100$ realizations of population distribution and movements (with learning) on a two-dimensional grid of size $N=50\times 50$. The errorbars are about the point sizes.}\label{SM2}
\end{figure}

We start with the results which are obtained by the simulated population distributions. Figures \ref{SM1} and \ref{SM2} display the average efficiency and the average relative entropies when only one parameter $g$ or $\Delta T_o$ changes. The KL divergence is divided by $\ln(2|E|)$ to be able to compare it for different city sizes. We see how much the two parameters contribute to the behaviour of the system efficiency. For comparison, the figures also show the results obtained with no learning, that is without any knowledge of the actual travel times in the previous cycles. Note that learning dose not necessarily increases the efficiency because the aim of learning here is just to find the shortest (time) path. Moreover, as the figures show, the learning process considerably changes the behaviour of the efficiency with the KL divergence. The relation with the relative entropy $\Delta S_T$, by contrast, does not qualitatively change by the learning process. The latter solely measures the changes in the size of the time intervals $\Delta T_d$ and $\Delta T_r$ which usually grow by increasing $g$ or $\Delta T_o$. On the other hand, the KL divergence is affected by both the size of the time intervals and the distribution of the arrival times in these intervals.

\begin{figure}
\includegraphics[width=16cm]{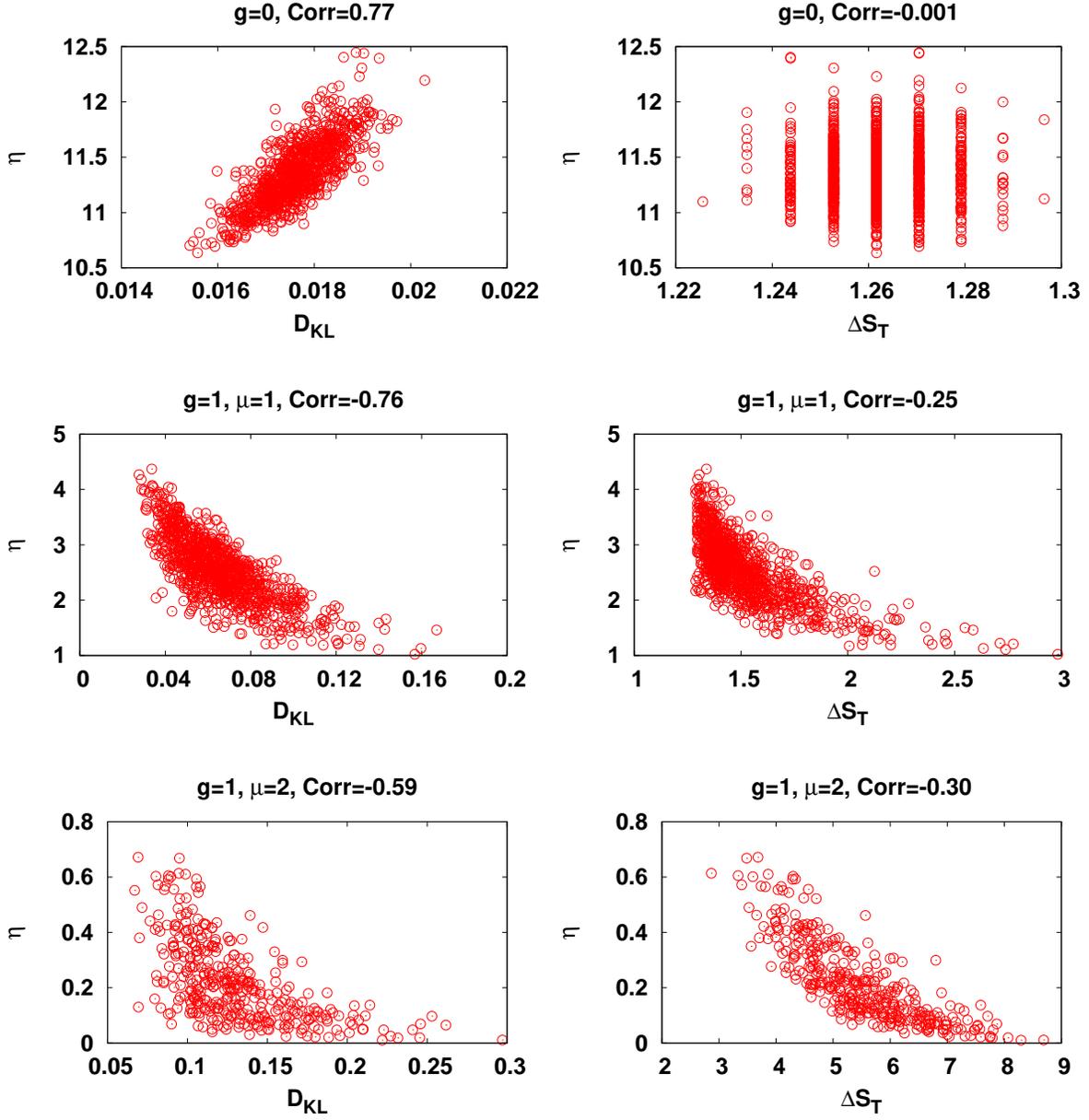} 
\caption{The efficiency $\eta$ vs the relative entropies $D_{KL}$ and $\Delta S_T$. Each point shows a realization of the simulated population distribution and movements (with learning) on a two-dimensional grid of size $N=50\times 50$. Here $\Delta t=1$ and $\Delta T_o=32$. The Pearson correlation coefficient shows the sign and magnitude of correlation between the two quantities.}\label{SM3}
\end{figure}

The behaviour of the efficiency with the relative entropies $D_{KL}(\overrightarrow{\mathbf{f}}||\overleftarrow{\mathbf{f}})$ and $\Delta S_T$ is shown in Fig. \ref{SM3} for some independent realizations of the population distribution. We observe a considerable negative correlation between the efficiency and the relative entropies, except for the case $g=0$, where the flows have no effect on the travel times. In this case, it is only the population distribution that determines the efficiency and the relative entropies. The positive correlation in this case is probably related to the fact that population distributions with closer ODs could result to smaller travel times $T$ but larger divergences $D_{KL}$ while the number of trips only changes slightly due to the small OD distances. Note that both the size of destination time interval $\Delta T_d$ and the distribution of arrival times in this interval affect the KL divergence whereas only the former is important for the relative entropy $\Delta S_T$. 

\begin{figure}
\includegraphics[width=16cm]{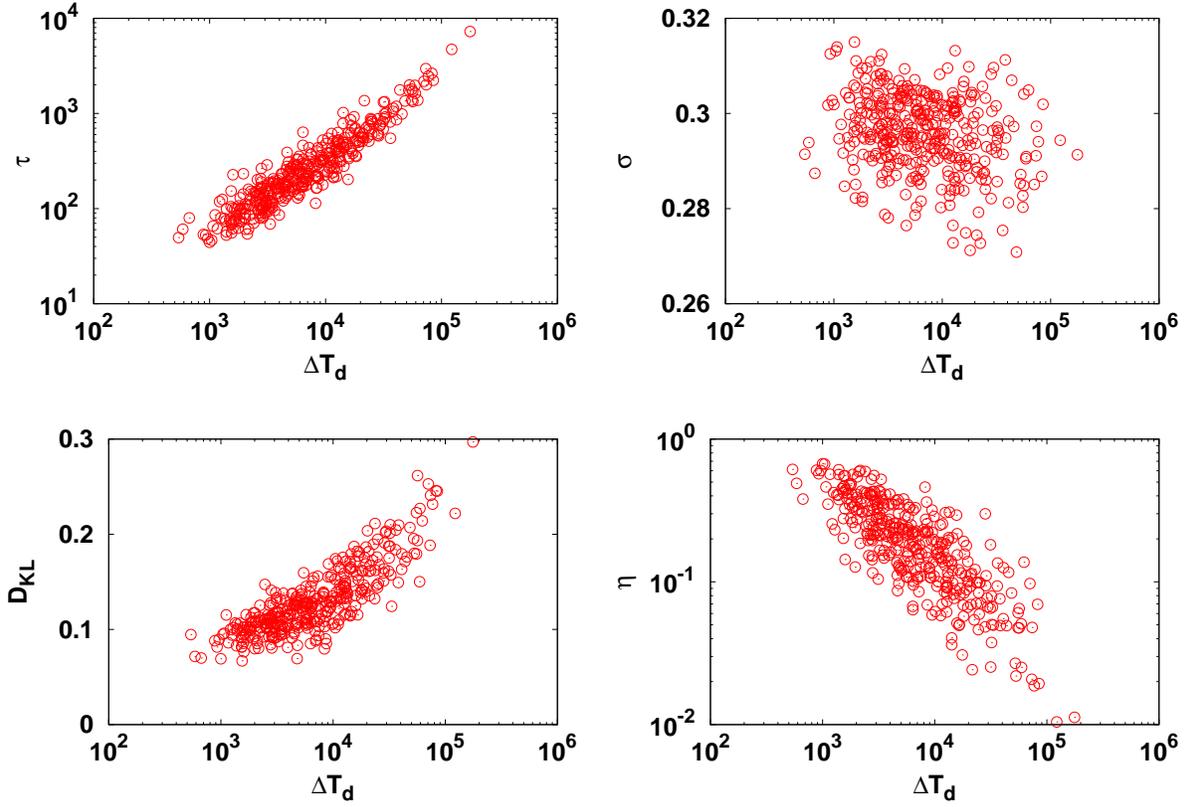} 
\caption{Dependence on the destination time interval $\Delta T_d$. The model parameters here are $\mu=2, g=1$ and $\Delta T_o=32$. Each point shows a realization of the simulated population distribution and movements (with learning) on a two-dimensional grid of size $N=50\times 50$. As before $\Delta t=1$.}\label{SM4}
\end{figure}

Dependence of the main quantities on $\Delta T_d$ is reported in Fig. \ref{SM4} for the case $(g=1,\mu=2)$. The destination time interval $\Delta T_d$ plays a central role in this study; the network structure and the impact of the forward flows on the travel times usually give rise to a large $\Delta T_d$ (larger than $\Delta T_o$). And the size of $\Delta T_d$ directly affects the divergence of the backward flows from the forward ones. Here both the forward and backward travel times $T_{OD},T_{DO}$ are expected to increase with $\Delta T_d$. Therefore, as the figure shows, the total travel time and $D_{KL}$ are positively correlated with $\Delta T_d$. On the other hand, we observe in Fig. \ref{SM4} that the total number of services $C$ is not very sensitive to $\Delta T_d$. This is true for both the forward and backward contributions $C_{OD}, C_{DO}$ as long as $\Delta T_o$ and $\Delta t$ are fixed (the latter here is set to one).

\subsection{Real population distributions}
\begin{figure}
\includegraphics[width=7cm]{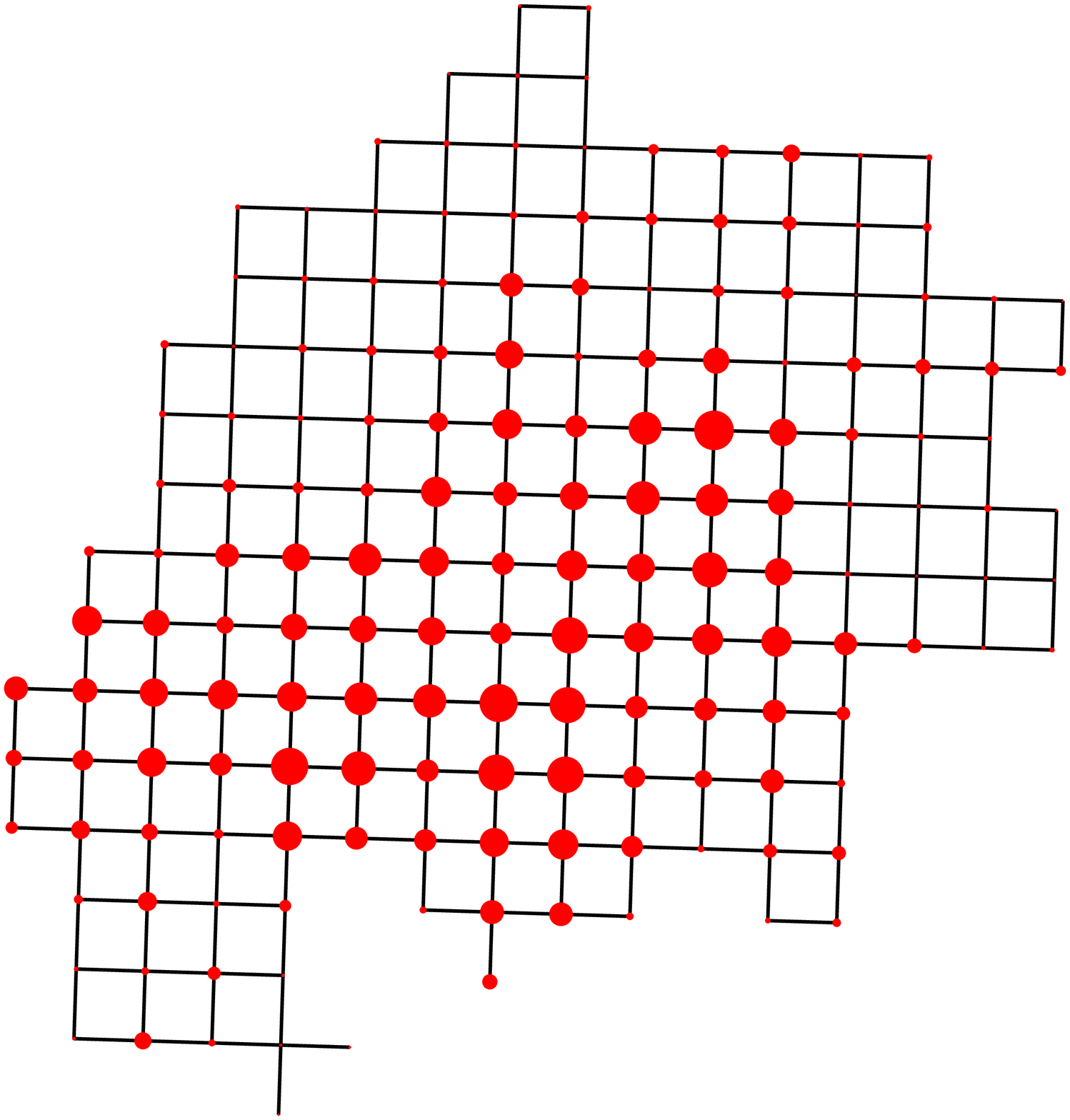} 
\includegraphics[width=9cm]{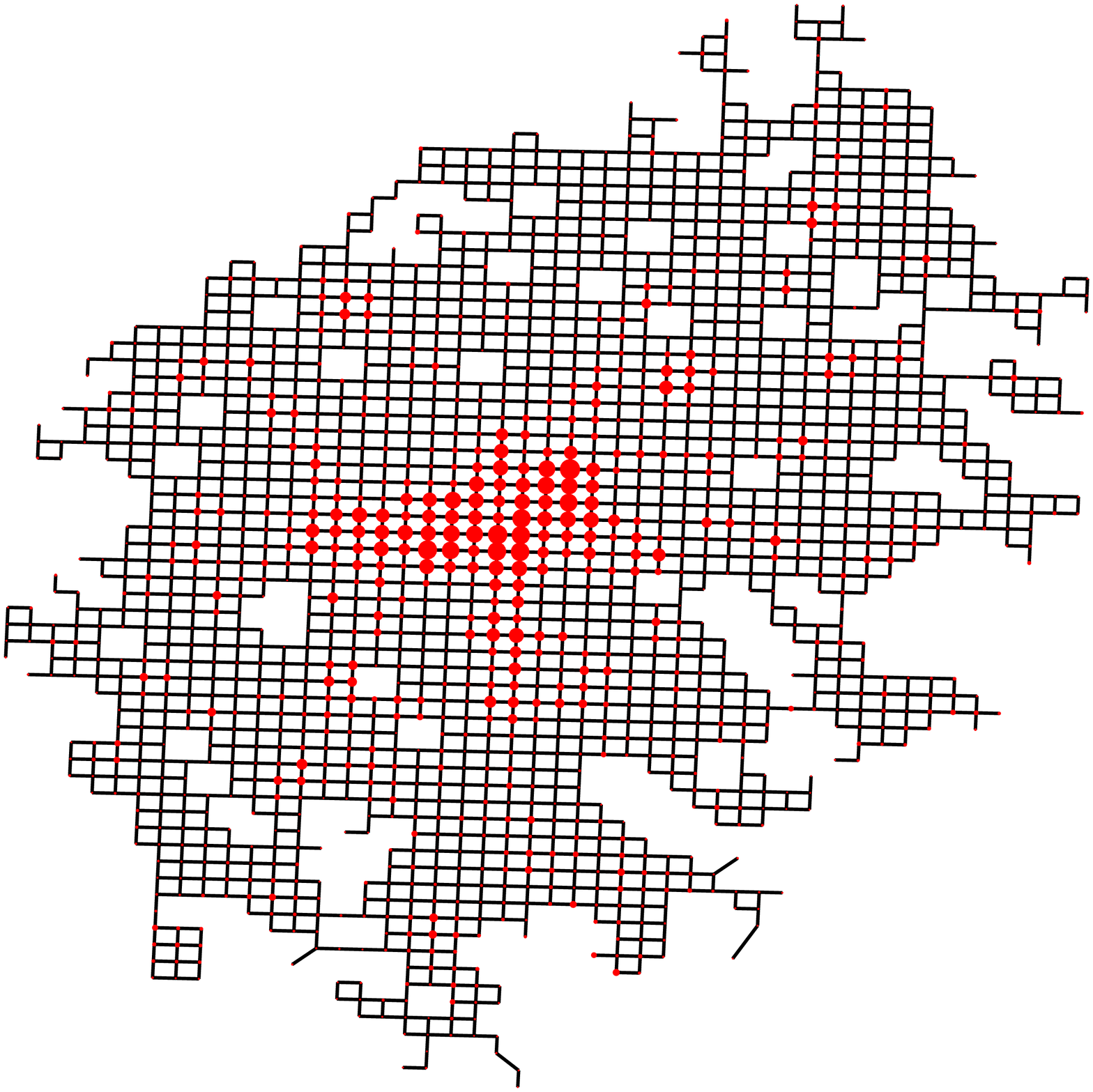} 
\caption{An example of the population distribution in a city (Turin). Left: the core part of the city ($N=169, M=935534$). Right: the core plus the commuting part ($N=1656, M=1772227$). The raw data are from \cite{popGridEurostat,sedacV4} and here are plotted with networkx(2.2).}\label{SM0}
\end{figure}

\begin{figure}
\includegraphics[width=16cm]{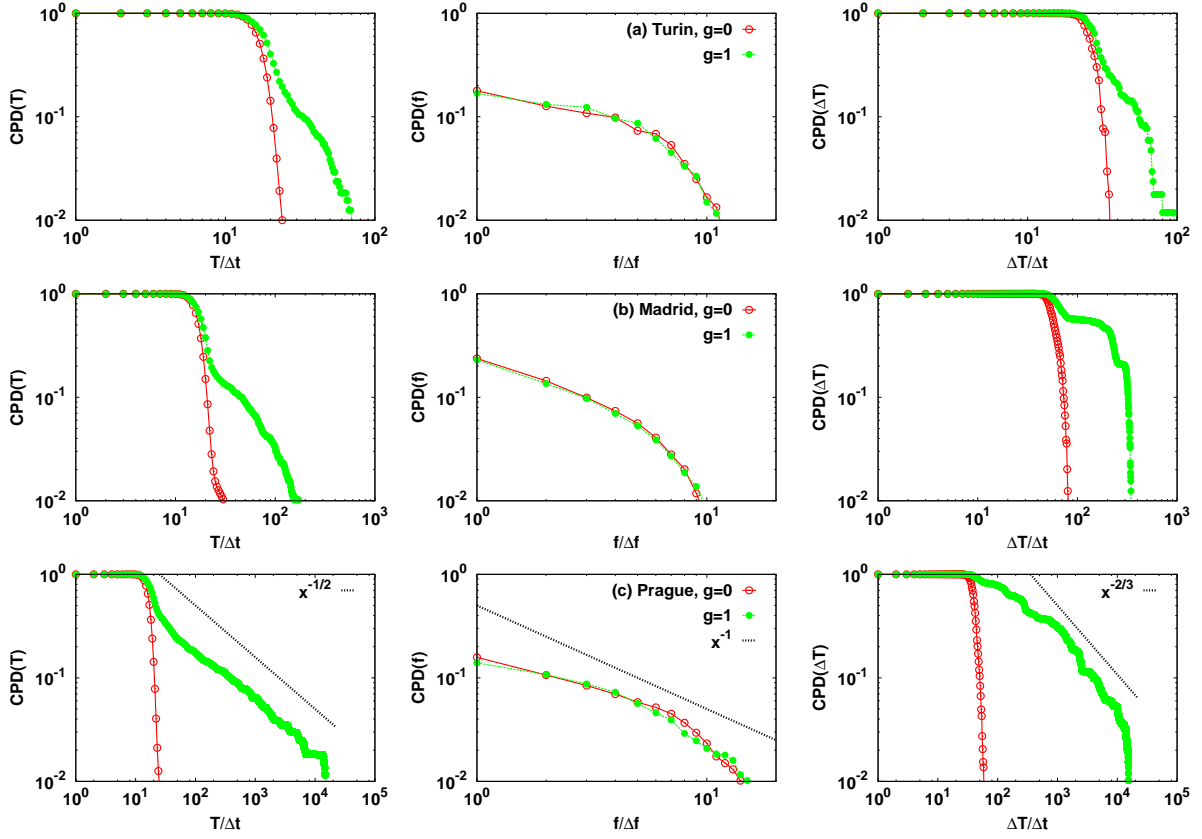} 
\caption{Cumulative probability distributions of simulated $T_{OD}$ (left), $f_{ab}$ (middle), and $\Delta T_d$ (right) in three cities. The data show distribution of $T_{OD}$ across the individuals $i$, $f_{ab}$ across the directed edges $(a,b)$, and $\Delta T_d$ across the network sites $a$. The data are obtained by numerical simulation of the movements after $20$ learning cycles using the population distributions of the cities. The model parameters are $\mu=3, g=1, \Delta T_o=32$ and $\Delta t=1$. For comparison we also report the results for the case $g=0$.}\label{SM5}
\end{figure}

Now, we report the results which are obtained by using the population distributions of $20$ real cities. Figure  \ref{SM0} displays such an example for the core and commuting parts of a real city. The cumulative distribution of the simulated OD times $T_{OD}$, the normalized flows $f_{ab}$, and the destination time intervals (across the sites) of three cities are displayed in Fig. \ref{SM5} for the core parts of the cities. Here, we observe a tendency to exhibit scale free behaviours by introducing the impact of the flows on the travel times. 

\begin{figure}
\includegraphics[width=14cm]{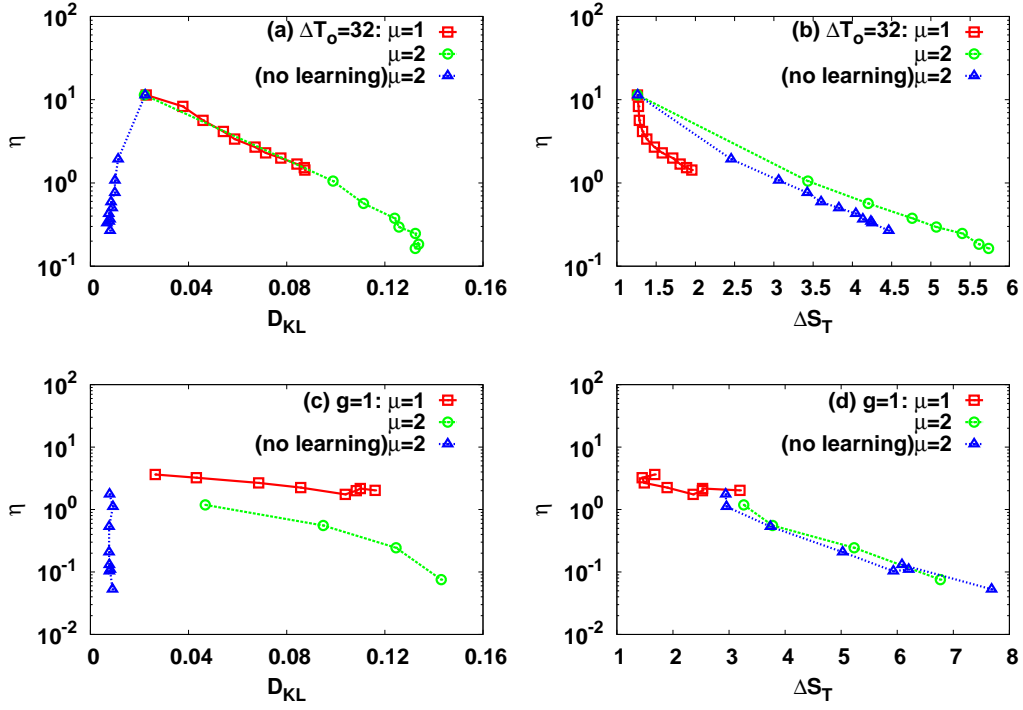} 
\caption{The average efficiency ($\eta$) vs the average relative entropies ($D_{KL}, \Delta S_T$). Top: the behaviour when only $g \in (0,2)$ changes with $\mu$ and $\Delta T_o$ fixed. Bottom: the behaviour when only $\Delta T_o \in (2^0,2^6)$ changes with $\mu$ and $g$ fixed. The average is taken over $100$ realizations of population distribution and movements (with learning) on a two-dimensional grid of size $N=50\times 50$. The errorbars are about the point sizes.}\label{SM1}
\end{figure}

\begin{figure}
\includegraphics[width=16cm]{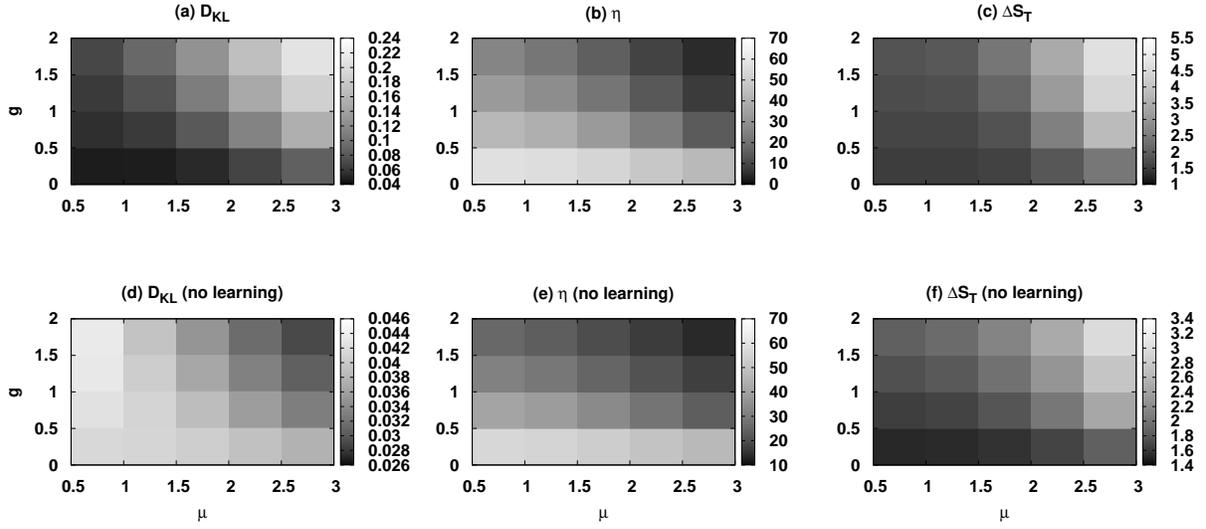} 
\caption{The average behaviour of the efficiency ($\eta$) and the relative entropies ($D_{KL}, \Delta S_T$) in $20$ cities (the core parts). The data are obtained by numerical simulation of the movements after $20$ learning cycles using the population distributions of the cities. The average is taken over the cities for $\Delta T_o=32$. For comparison we also report the results obtained without learning of the travel times (bottom panels).}\label{SM6}
\end{figure}

\begin{figure}
\includegraphics[width=14cm]{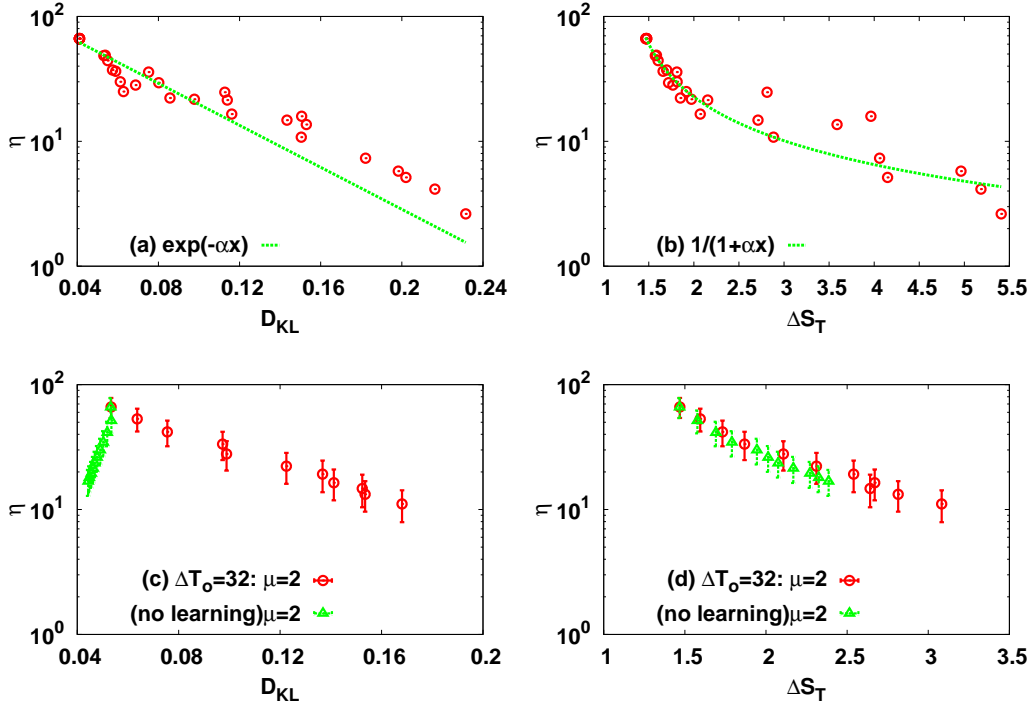} 
\caption{The average efficiency ($\eta$) vs the relative entropies ($D_{KL}, \Delta S_T$) for $20$ cities (the core parts). Top: the behaviour for different values of $g \in (0,2)$ and $\mu \in (0,3)$ for a given $\Delta T_o=32$ fixed. Bottom: the behaviour when only $g$ changes with $\mu$ and $\Delta T_o$ fixed.}\label{SM7}
\end{figure}

In Fig. \ref{SM6} we see how the average efficiency and the average relative entropies of these cities behave for various $\mu$ and $g$. Again, we observe that learning changes the sign of $(\eta, D_{KL})$ correlations whereas $\Delta S_T$ always shows a negative correlation with the efficiency. Figure \ref{SM7} displays the averages $\eta$, $D_{KL}$ and $\Delta S_T$ when the parameters $g$ and $\mu$ are changing for a fixed $\Delta T_o$. The data sets can well be described by an exponential relation $\eta \propto \exp(-\beta D_{KL})$, with exponent $\beta=19.3 \pm 1.8$. Similar behaviour is also observed with the simulated population distributions in Fig. \ref{SM1}.


\end{document}